\documentclass[a4paper,11pt]{article}

\usepackage{jheppub,bm,color}
\newcommand{\beq}{\begin{eqnarray}}
\newcommand{\eeq}{\end{eqnarray}}
\renewcommand\d{\partial}

\newcommand{\SU}{\text{SU}}
\newcommand{\U}{\text{U}}

\renewcommand\l{\lambda}

\renewcommand\t{\tau}

\renewcommand\o{\omega}

\newcommand\m{\mu}
\newcommand\n{\nu}
\newcommand\x{\xi}
\newcommand\p{\pi}
\newcommand\h{\theta}

\newcommand\f{\phi}

\renewcommand\S{\Sigma}
\renewcommand\O{\Omega}

\renewcommand{\vec}[1]{\bm{#1}}
\newcommand\lb{\left(}
\newcommand\rb{\right)}
\newcommand\ls{\left[}
\newcommand\rs{\right]}

\newcommand{\non}{\nonumber\\}
\newcommand\pt{\partial}

\newcommand{\Tr}{{\rm Tr}}

\renewcommand{\part}{{\rm part}}

\begin{document}

\title{Anomalous effects of dense matter under rotation}

\author[a,b]{Xu-Guang Huang,}
\author[c]{Kentaro Nishimura}
\author[c]{and Naoki Yamamoto}

\affiliation[a]{Physics Department and Center for Particle Physics and Field Theory, Fudan University, \\ Shanghai 200433, China}
\affiliation[b]{Key Laboratory of Nuclear Physics and Ion-beam Application (MOE), Fudan University, \\ Shanghai 200433, China}
\affiliation[c]{Department of Physics, Keio University, Yokohama 223-8522, Japan}

\emailAdd{huangxuguang@fudan.edu.cn}
\emailAdd{nishiken@a6.keio.jp}
\emailAdd{nyama@rk.phys.keio.ac.jp}

\abstract
{We study the anomaly induced effects of dense baryonic matter under rotation.
We derive the anomalous terms that account for the chiral vortical effect in the low-energy
effective theory for light Nambu-Goldstone modes.
The anomalous terms lead to new physical consequences, such as the anomalous Hall energy
current and spontaneous generation of angular momentum in a magnetic field
(or spontaneous magnetization by rotation). In particular, we show that, due to
the presence of such anomalous terms, the ground state of the quantum chromodynamics (QCD) under sufficiently fast rotation
becomes the ``chiral soliton lattice" of neutral pions that has lower energy than the QCD
vacuum and nuclear matter. We briefly discuss the possible realization of the chiral soliton lattice
induced by a fast rotation in noncentral heavy ion collisions.}
\maketitle

\section{Introduction}
Understanding of matter under extreme conditions is an important question in quantum
chromodynamics (QCD). Among others, QCD matter under fast rotation has attracted
increasing attention. Theoretically, it has been expected that noncentral heavy ion collisions may
produce quark-gluon fluids with very large angular momentum which may lead to spin polarization
of hadrons due to the spin-orbit coupling \cite{Liang:2004ph, Becattini:2007sr, Huang:2011ru,
Csernai:2013bqa, Becattini:2013vja, Becattini:2015ska, Jiang:2016woz, Pang:2016igs, Deng:2016gyh}.
In fact, recent heavy ion collision experiments at the Relativistic Heavy Ion Collider (RHIC) reported
an alignment between the global angular momentum of a noncentral collision and the spin of emitted
$\Lambda$ hyperons, indicating the generation of the largest vorticity of fluids observed so far
\cite{STAR:2017ckg}. In such relativistic fluids, the anomalous transport phenomena along the
vorticity, called the chiral vortical effect (CVE) \cite{Vilenkin:1979ui, Vilenkin:1980zv, Erdmenger:2008rm, Banerjee:2008th,
Son:2009tf, Landsteiner:2011cp, Landsteiner:2012kd, Landsteiner:2016led}, and a resulting collective mode,
called the chiral vortical wave \cite{Jiang:2015cva}, may be realized \cite{Kharzeev:2015znc}.
The phases of QCD matter under global rotation has also been studied based on effective models
\cite{Chen:2015hfc, Jiang:2016wvv, Ebihara:2016fwa, Chernodub:2016kxh, Chernodub:2017ref}
and on the lattice \cite{Yamamoto:2013zwa}.

In this paper, we study the generic and model-independent aspects of low-temperature and dense
baryonic matter under rotation that are tied to the quantum anomalies. Due to the exactness
of (part of) the transport coefficients of the CVE, one can provide stringent constraints
on the low-energy effective theory of QCD. Previously, anomalous terms in dense matter
with background electromagnetic fields were studied \cite{Son:2004tq, Son:2007ny}.
Here, we derive new anomalous terms that account for the CVE in the effective theory for
light Nambu-Goldstone (NG) modes; see eq.~(\ref{L_new}) for the main result. This may be
viewed as the ``anomaly matching" for CVE. We argue that the presence of these anomalous
terms leads to dramatic new physical consequences, such as the anomalous Hall energy current
and spontaneous generation of angular momentum in a magnetic field or spontaneous
magnetization by rotation. In particular, we show that the ground state of QCD under
sufficiently fast rotation becomes the ``chiral soliton lattice" of neutral pions that has lower
energy than the QCD vacuum and nuclear matter, similarly to the situation of QCD in a strong
magnetic field \cite{Son:2007ny, Brauner:2016pko}. As we will argue, such an exotic state under
fast rotation may potentially be realized in future low-energy heavy ion collision experiments.

This paper is organized as follows. In section~\ref{sec:CVE}, we derive the anomaly induced terms
in a rotating matter. In section~\ref{sec:nuclear}, we consider two-flavor QCD at finite baryon and isospin
chemical potentials under rotation and study its ground state. In section~\ref{sec:physics}, we consider
the physical consequences of the anomalous terms. Section~\ref{sec:discussion} is devoted to discussions.

Throughout this paper, we use the sign convention $(+,-,-,-)$ for the metric tensor.

\section{Anomaly matching for chiral vortical effect}
\label{sec:CVE}
We consider QCD with $N_{\rm f}$ flavors at finite chemical potential and temperature $T$
under a global rotation ${\bm \Omega}$.%
\footnote{We assume the size of the system satisfies $R < 1/|{\bm \Omega}|$ such that the
velocity of the boundary is less than the speed of light.} In this section, we consider massless quarks
(chiral limit). Generically, there can be more than one chemical potentials,
$\mu_a$, coupled to conserved charges $\bar q \gamma^0 \tau_a q$, where $\tau_{a}$
($a=0,1,2,\cdots,N_{\rm f}^2-1$) are the  $\U(N_{\rm f})$ generators with the normalization $\Tr(\tau_a \tau_b) =2 \delta_{ab}$.
In relativistic matter, the axial current ${\bm j}_a^5 = \langle \bar q {\bm \gamma} \gamma^5 \tau_a q \rangle$
can be generated at finite chemical potential $\mu_a$ and finite temperature $T$ along the rotation
${\bm \Omega}$, which is called the CVE \cite{Vilenkin:1979ui, Vilenkin:1980zv, Erdmenger:2008rm, Banerjee:2008th, Son:2009tf, Landsteiner:2011cp, Landsteiner:2012kd, Landsteiner:2016led}.
Without a chiral chemical potential, the expression of the CVE is given by
\cite{Landsteiner:2011cp, Landsteiner:2012kd, Landsteiner:2016led}
\begin{gather}
{\bm j}_a^5 = N_{\rm c} \left(d_{abc} \frac{\mu_b \mu_c}{2\pi^2} + b_a \frac{T^2}{6} \right) {\bm \Omega}\,,
\nonumber \\
d_{abc} = \frac{1}{2}\Tr \left[\tau_a \{\tau_b, \tau_c \} \right], \qquad b_a = \Tr(\tau_a),
\label{CVE}
\end{gather}
where $N_{\rm c}$ is the number of colors.

It is known that the coefficients of the $\mu$ and $T$ dependent terms in eq.~(\ref{CVE}) are
respectively related to the chiral anomaly and the gauge-gravitational anomaly
\cite{Landsteiner:2011cp, Landsteiner:2012kd, Landsteiner:2016led}, and that the coefficient of the
$\mu$-dependent term does not receive any renormalization
(while the $T$-dependent term can when it couples to dynamical gauge fields \cite{Golkar:2012kb, Hou:2012xg}).
This nonrenormalization is a consequence of the topological nature of the underlying quantum field theory,
similarly to that of the chiral anomaly. Hence, one expects that the CVE should appear independently of
the energy scale involved and the absence/presence of spontaneous symmetry breaking of the system.
In particular, it should be reproduced in the framework of the low-energy effective theory for
NG modes---the chiral perturbation theory (ChPT).
This is analogous to the fact that the chiral anomaly \cite{Wess:1971yu, Witten:1983tw} and the chiral magnetic effect
\cite{Fukushima:2012fg, Cohen} can be reproduced in the ChPT by taking into account the
Wess-Zumino-Witten term appropriately. In order to look for new terms responsible for the CVE in the
ChPT under rotation, we will rather \emph{postulate} such a matching without a rigorous proof in this paper.

Consider a \emph{local} chiral rotation,
\beq
\label{axial}
q \rightarrow {\rm e}^{-i \theta_a \gamma_5 \tau_a} q,
\eeq
with $\theta_a = \theta_a(x^{\mu})$.
Under this transformation, the QCD action changes by the amount,
\beq
\label{S_QCD}
\delta S_{\rm QCD} = \int {\rm d}^4 x (\d_{\mu} \theta_a) j_a^{5 \mu},
\eeq
where $j_a^{5 \mu} = \bar q \gamma^{\mu} \gamma^5 \tau_a q$.
Here we used $\d_{\mu} j_a^{5 \mu} = 0$ in the chiral limit.

At low energy, on the other hand, the axial current $j^{5\mu}_a$ is carried by a NG mode $\pi_a$
associated with chiral symmetry breaking. In the low-energy effective theory, this NG field is
expressed by the nonlinear representation, $\Sigma \sim \exp \left(i \pi_a \tau_a/f_{\pi} \right)$ with $f_{\pi}$ the decay constant (which we take to be the same for all the NG modes under the
$\SU(N_{\rm f})$ vector symmetry).
Recalling that $\pi_a$ transforms under the chiral rotation~(\ref{axial}) as
\beq
\pi_a \rightarrow \pi_a + 2 f_{\pi} \theta_a,
\eeq
the change of the action associated with the rotation (\ref{axial}) can be expressed as
\beq
\label{S_EFT}
\delta S_{\rm EFT} = \int {\rm d}^4 x (\d_{\mu} \alpha_a) j_a^{5 \mu},
\eeq
where $\alpha_a = \pi_a/(2f_{\pi})$.

Now we require the matching condition $\delta S_{\rm QCD} = \delta S_{\rm EFT}$ between
eqs.~(\ref{S_QCD}) and (\ref{S_EFT}).
This condition leads to the following new term in the low-energy effective theory:
\beq
\label{L_new}
{\cal L}_{\rm EFT} =  \frac{N_{\rm c}}{2f_{\pi}}
\left(\frac{d_{abc}}{2\pi^2} \mu_b \mu_c
+ \frac{b_a}{6} T^2 \right) {\bm \nabla} \pi_a  \cdot {\bm \Omega}\,,
\eeq
or
\beq
\label{L_new2}
{\cal L}_{\rm EFT} =-  \frac{N_{\rm c} }{2f_{\pi}} \pi_a
\left(\frac{d_{abc}}{\pi^2} \mu_b {\bm \nabla} \mu_c
+ \frac{b_a}{3} T {\bm \nabla} T \right) \cdot {\bm \Omega}\,,
\eeq
up to total derivatives.

If $\pi_a/(2f_{\pi})$ is regarded as the $\theta$ term, the $T$-dependent term
$\sim\theta T {\bm \nabla} T \cdot {\bm \Omega}$ in eq.~(\ref{L_new2}) takes exactly the same form
as the thermodynamic analog of the Chern-Simons term that appears in 3D topological superfluids
\cite{Nomura:2011hn} including the superfluid $^3$He \cite{Volovik:1999wx}.
However, the $\mu$-dependent term has eluded in such a context to the best of our knowledge.

As we explained above, the coefficient of the $\mu$-dependent part should not receive any renormalization
and should be exact. On the other hand, the $T$-dependent part may have possible renormalization.
In the following, we will focus on the cases where this exact $\mu$-dependent part is relevant.
(We will briefly comment on the case where the $T$-dependent part is also relevant in appendix~\ref{sec:CSC}.)

\section{QCD with both $\mu_{\rm B}$ and $\mu_{\rm I}$ under rotation}
\label{sec:nuclear}
Our discussion so far has been general. In the following, we will consider a more specific case:
two-flavor QCD at finite baryon and isospin chemical potentials, $\mu_{\rm B}$ and $\mu_{\rm I}$,
under rotation. In this case, the $\mu$-dependent anomalous term (\ref{L_new}) appears for the
neutral $\pi_0$ meson and it is written as%
\footnote{The normalization of $\mu_{\rm I}$ here is chosen such that it matches that of the
previous literature including ref.~\cite{Son:2000xc}.}
\beq
\label{L_pi}
{\cal L}_{\rm anom} =  \frac{\mu_{\rm B} \mu_{\rm I}}{2\pi^2 f_{\pi}} {\bm \nabla} \pi_0 \cdot {\bm \Omega}\,,
\eeq
Note that the $T$-dependent term is absent for $\pi_0$ since $b_3 = 0$.

This term accounts for the angular momentum density, baryon and isospin number densities
in the presence of finite $\langle {\bm \nabla} \pi_0 \rangle$ as
\beq
\label{j}
{\bm j} = \frac{\mu_{\rm B} \mu_{\rm I}}{2\pi^2 f_{\pi}} \langle {\bm \nabla} \pi_0 \rangle\,, \quad
n_{\rm B} = \frac{\mu_{\rm I}}{2\pi^2 f_{\pi}} \langle {\bm \nabla} \pi_0 \rangle \cdot {\bm \Omega}\,, \quad
n_{\rm I} = \frac{\mu_{\rm B}}{2\pi^2 f_{\pi}} \langle {\bm \nabla} \pi_0 \rangle \cdot {\bm \Omega}\,,
\eeq
independent of the origin of $\langle {\bm \nabla} \pi_0 \rangle$.
(Here and below, ${\bm j}$ is not to be confused with the vector current.)
Without loss of generality, we take ${\bm \Omega}=\Omega \hat{\bm z}$ and use
a cylindrical coordinate system $(r, \theta, z)$ below in this section.

\subsection{Chiral limit}
We first consider the simplest case that the quarks are massless.
The Hamiltonian density for the neutral $\pi_0$ at finite $\mu_{\rm B}$ and $\mu_{\rm I}$
under rotation is
\beq
{\cal H} = \frac{1}{2} \left[(\d_r \pi_0)^2 + \frac{1-(\Omega r)^2}{r^2} (\d_{\theta} \pi_0)^2 + (\d_z \pi_0)^2 \right] - \frac{\mu_{\rm B} \mu_{\rm I}}{2\pi^2 f_{\pi}} \Omega \d_z \pi_0\,,
\eeq
where we also added the nonanomalous contribution of the global rotation to the kinetic term.
This takes the minimum value,
\beq
\langle {\cal H} \rangle = - \frac{\mu_{\rm B}^2 \mu_{\rm I}^2}{8\pi^4 f_{\pi}^2} \Omega^2\,,
\eeq
when $\langle \d_ r \pi_0 \rangle = \langle \d_{\theta} \pi_0 \rangle = 0$ (for $\Omega r <1$) and
\beq
\label{profile}
\langle \d_z \pi_0 \rangle = \frac{\mu_{\rm B} \mu_{\rm I}}{2\pi^2 f_{\pi}} \Omega \,, \qquad
{\rm or} \qquad
\langle \pi_0 \rangle = \frac{\mu_{\rm B} \mu_{\rm I}}{2\pi^2 f_{\pi}} \Omega z + {\rm const}.
\eeq
The angular momentum density, baryon and isospin number densities of this state are found,
using (\ref{j}), as
\beq
j = \frac{\mu_{\rm B}^2 \mu_{\rm I}^2}{4\pi^4 f_{\pi}^2} \Omega \,, \quad
n_{\rm B} = \frac{\mu_{\rm B} \mu_{\rm I}^2}{4\pi^4 f_{\pi}^2} \Omega^2\,, \quad
n_{\rm I} = \frac{\mu_{\rm B}^2 \mu_{\rm I}}{4\pi^4 f_{\pi}^2} \Omega^2\,.
\eeq
In particular, note that $n_{\rm B} \gg n_{\rm I}$ when $\mu_{\rm B} \ll \mu_{\rm I}$, which indicates
that the rotation can induce a very large baryon (isospin) susceptibility at high $\mu_{\rm I}$ ($\mu_{\rm B}$).

\subsection{Finite quark masses: rotation-induced chiral soliton lattice}
\label{sec:CSL}
In the presence of nonzero quark masses, $\pi_0$ forms a ``chiral soliton lattice" (CSL) similarly to
the one found in an external magnetic field ${\bm B}$ \cite{Brauner:2016pko}. The Hamiltonian of the
$\pi_0$ sector with a nonzero pion mass $m_{\pi}$ is
\beq
\label{H}
\! \! {\cal H} = \frac{f_{\pi}^2}{2} \! \left[(\d_r \phi)^2 + \frac{1-(\Omega r)^2}{r^2} (\d_{\theta} \phi)^2 + (\d_z \phi)^2 \right] \!
+ m_{\pi}^2 f_{\pi}^2 (1- \cos \phi) - \frac{\mu_{\rm B} \mu_{\rm I}}{2\pi^2} \Omega \d_z \phi \,,
\eeq
where we introduced $\phi \equiv \pi^0/f_{\pi}$. Similarly to the case in the chiral limit above, the $r$
and $\theta$ components of the Hamiltonian are minimized when
$\langle \d_ r \pi_0 \rangle = \langle \d_{\theta} \pi_0 \rangle = 0$ for $\Omega r <1$.
The remaining $z$ component of the Hamiltonian is mathematically equivalent to that at finite $\mu_{\rm B}$
in an external magnetic field in ref.~\cite{Brauner:2016pko} by the replacement ${\bm B} \rightarrow 2\mu_{\rm I} {\bm \Omega}$.
Hence, the ground state (and excited states) of this system can be found using the results already obtained
in ref.~\cite{Brauner:2016pko}, which we briefly summarize below.

We will focus on the $z$ component of the Hamiltonian. Then, the equation of motion is
\beq
\d_z^2 \phi = m_{\pi}^2 \sin \phi,
\eeq
which can be solved analytically using the Jacobi elliptic function as
\beq
\cos \frac{\phi(\bar z)}{2} = {\rm sn}(\bar z, k)\,,
\eeq
where $\bar z \equiv z m_{\pi}/k$ is a dimensionless coordinate and $k$ ($0 \leq k \leq 1$) is
the elliptic modulus. This solution suggests the lattice structure with the period,
\beq
\ell = \frac{2k K(k)}{m_{\pi}}\,,
\eeq
where $K(k)$ is the complete elliptic integral of the first kind.

Each unit cell of the lattice carries an angular momentum, baryon and isospin charges
per unit area in the $xy$ plane,%
\footnote{We denote the area in the $xy$ plane by $A$ instead of $S$, as the latter
will be used for the entropy later.}
\beq
\frac{J_z}{A} = \frac{\mu_{\rm B} \mu_{\rm I}}{\pi}\,, \quad
\frac{N_{\rm B}}{A} = \frac{\mu_{\rm I} \Omega}{\pi} \,, \quad
\frac{N_{\rm I}}{A} = \frac{\mu_{\rm B} \Omega}{\pi} \,.
\eeq
These quantities are topological charges in the sense that they do not depend on the
details of $\phi(z)$, but only on their boundary values of the lattice.
Interestingly, the last two relations represent the cross-correlated responses induced by rotation:
$\mu_{\rm I}$ induces $N_{\rm B}$ and $\mu_{\rm B}$ induces $N_{\rm I}$.

By minimizing the total energy of the system at fixed volume with respect to $k$, one obtains
the optimal $k$ for given $\mu_{\rm B}$, $\mu_{\rm I}$, and $\Omega$ as
\beq
\label{condition}
\frac{E(k)}{k}= \frac{\mu_{\rm B} |\mu_{\rm I}| \Omega}{8\pi m_{\pi} f_{\pi}^2}\,,
\eeq
where $E(k)$ is the complete elliptic integral of the second kind.
Because $E(k)/k \geq 1$, the condition for the existence of the CSL solution is given by
\beq
\label{Omega_CSL}
|\Omega| \geq \Omega_{\rm CSL} \equiv \frac{8\pi m_{\pi} f_{\pi}^2}{\mu_{\rm B} |\mu_{\rm I}|}\,.
\eeq
One can also show that, when this condition is satisfied, the CSL has a lower energy than
the QCD vacuum and nuclear matter. In fact, the total energy of the system at a fixed volume $V$
under the condition (\ref{condition}) is given by (see eq.~(4.8) of ref.~\cite{Brauner:2016pko}):
\beq
\label{E_tot}
\frac{{\cal E}_{\rm tot}}{V} = 2m_{\pi}^2 f_{\pi}^2\left(1-\frac{1}{k^2} \right)<0\,.
\eeq
Therefore, the CSL is the genuine QCD ground state for $|\Omega| \geq \Omega_{\rm CSL}$
(as long as Bose-Einstein condensation of charged pions does not occur \cite{Brauner:2016pko};
see appendix \ref{sec:CPC} for discussion about the charged pion condensation under rotation).
Note, in particular, that $\Omega_{\rm CSL} = 0$ in the chiral limit where $m_{\pi} = 0$. In this case,
nuclear matter would be unstable under an infinitesimally small rotation.

\section{Physical consequences}
\label{sec:physics}
In this section, we study two physical consequences of the new term (\ref{L_pi}): anomalous Hall
energy current and cross-correlated responses between ${\bm \Omega}$ and ${\bm B}$.

\subsection{Anomalous Hall energy current}
\label{sec:Hall}
A nontrivial consequence of (\ref{L_pi}) is the generation of the anomalous Hall energy current:
when $\mu_{\rm B}$ is inhomogeneous under a rotation ${\bm \Omega} = \Omega \hat{\bm r}$,
an energy current is induced in the direction of ${\bm \nabla}\mu_{\rm B} \times \hat {\bm r}$.
The Hall coefficient $c_{\rm H}$ is defined as
\beq
\label{def:c_H}
T^{0i} = c_{\rm H} \epsilon^{ijk} (\d_j \mu_{\rm B}) \hat r^k.
\eeq

To derive $c_{\rm H}$ explicitly, we take into account the effects of the global rotation ${\bm \Omega}$
by considering a gravitational field whose metric is given by
\beq
{\rm d} s^2 = {\rm d} t^2 + 2 g_{0i} {\rm d}t {\rm d}x^i - ({\rm d}x^i)^2,
\eeq
with $g_{0i}$ satisfying $\Omega^i = \frac{1}{2}\epsilon^{ijk}\d_j g_{0k}$.
Then, we can compute the energy current by $T^{0i} = 2 (\delta S_{\rm anom}/\delta g_{0i})$ for
\beq
S_{\rm anom} = \frac{\mu_{\rm I}}{2\pi^2 f_{\pi}} \int {\rm d}^4 x \ \mu_{\rm B} {\bm \nabla} \pi_0 \cdot {\bm \Omega}\,,
\eeq
with $\mu_{\rm B} = \mu_{\rm B}({\bm x})$, to obtain
\beq
\label{c_H}
c_{\rm H} = \frac{\mu_{\rm I}}{2\pi^2 f_{\pi}} \left(\langle \d_z \pi_0 \rangle
+ \frac{\d \langle \d_z \pi_0 \rangle}{\d \Omega} \Omega \right)\,,
\eeq
where we again take the global rotation in the $z$ direction ${\bm \Omega}= \Omega \hat {\bm z}$.
Note that this is valid independently of quark masses. Using (\ref{j}), the Hall coefficient above
can also be expressed as the following St\v{r}eda-type formula:
\beq
\label{Streda}
c_{\rm H} = \left(\frac{\d n_{\rm B}}{\d \Omega} \right)_{\! \! \mu_{\rm B}}\,.
\eeq

In the chiral limit where the profile of $\langle \d_z \pi_0 \rangle$ is given by (\ref{profile}),
the Hall coefficient~(\ref{c_H}) becomes
\beq
c_{\rm H} = 2\mu_{\rm B}\Omega \left(\frac{\mu_{\rm I}}{2\pi^2 f_{\pi}}\right)^{\! 2} \,.
\eeq

In the presence of finite quark masses, the profile of $\langle \d_z \pi_0 \rangle$ is given by the
inhomogeneous CSL as we have seen in section~\ref{sec:CSL}. In this case, it is convenient to
define the average baryon number density,
\beq
\label{n_bar}
\bar n_{\rm B} \equiv - \frac{1}{V} \frac{\d {\cal E}_{\rm tot}}{\d \mu_{\rm B}}
= -\frac{4 m_{\pi}^2 f_{\pi}^2}{k^3}\frac{\d k}{\d \mu_{\rm B}}\,,
\eeq
where we used eq.~(\ref{E_tot}) in the last equation. By using the relation,
\beq
\frac{\d k}{\d \mu_{\rm B}} = - \frac{k^2 \mu_{\rm I} \Omega}{8\pi m_{\pi} f_{\pi}^2 K(k)}\,,
\eeq
that follows from the condition~(\ref{condition}), we have
\beq
\bar n_{\rm B} = \frac{m_{\pi} \mu_{\rm I} \Omega}{2\pi k K(k)}\,.
\eeq
Substituting it to the St\v{r}eda-type formula (\ref{Streda}) and using the relation,
\beq
\label{dkdOmega}
\frac{\d k}{\d \Omega} = - \frac{k^2 \mu_{\rm B} \mu_{\rm I}}{8\pi m_{\pi} f_{\pi}^2 K(k)}\,,
\eeq
that also follows from the condition~(\ref{condition}), we obtain the average Hall coefficient:
\beq
\bar c_{\rm H} = \frac{m_{\pi} \mu_{\rm I}}{2\pi k^2 K(k)^2} + \frac{\mu_{\rm B} \mu_{\rm I}^2 E(k) \Omega}{16 \pi^2 m_{\pi} f_{\pi}^2 (1-k^2)K(k)^3 }\,.
\eeq

\subsection{Cross-correlated responses between rotation and magnetic field}
\label{sec:response}
Another interesting consequence of the anomalous term (\ref{L_pi}) is the
cross-correlated responses between ${\bm \Omega}$ and ${\bm B}$.
Remember first that, in the presence of the magnetic field ${\bm B}$, there is an additional
anomalous term proportional to ${\bm \nabla} \pi_0 \cdot {\bm B}$ \cite{Son:2007ny}
related to the chiral magnetic effect.

Let us first consider the chiral limit, where the Hamiltonian under rotation
{\it and} in a magnetic field is
\beq
{\cal H} = \frac{1}{2} ({\bm \nabla}\pi_0)^2 - \frac{\mu_{\rm B}}{4\pi^2 f_{\pi}}
{\bm \nabla} \pi_0 \cdot (2\mu_{\rm I}{\bm \Omega} + {\bm B}) \,,
\eeq
where we assumed a sufficiently small global rotation $|{\bm \Omega}| r \ll 1$ and ignored its
contribution to the kinetic term.
Integrating out $\pi^0$, we get an effective Hamiltonian of the classical fields ${\bm \Omega}$
and ${\bm B}$, in which the mixing term is given by
\beq
{\cal H}_{\rm mix} = -\frac{\mu_{\rm B}^2 \mu_{\rm I}}{8\pi^4 f_{\pi}^2} {\bm \Omega} \cdot {\bm B}\,.
\eeq
This term induces a {\it spontaneous} angular momentum density in a magnetic field
or a {\it spontaneous} magnetization density due to rotation,
\begin{gather}
{\bm j} = \chi_{jB} {\bm B}\,, \qquad
{\bm m} =  \chi_{m\Omega} {\bm \Omega}\,, \\
\chi_{jB} = \chi_{m \Omega} = \frac{\mu_{\rm B}^2 \mu_{\rm I}}{8\pi^4 f_{\pi}^2}\,.
\end{gather}
This provides a new example of the cross-correlated responses between ${\bm \Omega}$ and ${\bm B}$.

For finite quark masses, the angular momentum density is computed using
eqs.~(\ref{E_tot}) and (\ref{dkdOmega}) as
\beq
j \equiv - \frac{1}{V}\frac{\d {\cal E}_{\rm tot}}{\d \Omega} = \frac{m_{\pi} \mu_{\rm B} \mu_{\rm I}}{2 \pi k K(k)}\,.
\eeq
The susceptibility is then
\beq
\label{chi_jB}
\chi_{jB} = \frac{\mu_{\rm B}^2 \mu_{\rm I} E(k)}{32 \pi^2 f_{\pi}^2 (1-k^2)K(k)^3}\,,
\eeq
where we used the relation,
\beq
\frac{\d k}{\d B} = -\frac{k^2 \mu_{\rm B}}{16 \pi m_{\pi} f_{\pi}^2 K(k)}\,,
\eeq
which follows by taking the derivative of eq.~(4.5) in ref.~\cite{Brauner:2016pko} with respect to $B \equiv |{\bm B}|$.
Similarly, we obtain the magnetization,
\beq
m \equiv - \frac{1}{V}\frac{\d {\cal E}_{\rm tot}}{\d B} = \frac{m_{\pi} \mu_{\rm B}}{4\pi k K(k)}\,,
\eeq
and the susceptibility,
\beq
\label{chi_mOmega}
\chi_{m \Omega} = \frac{\mu_{\rm B}^2 \mu_{\rm I} E(k)}{32 \pi^2 f_{\pi}^2 (1-k^2)K(k)^3}\,,
\eeq
which is equal to eq.~(\ref{chi_jB}).

Note here that the relation $\chi_{jB} = \chi_{m \Omega}$ is a consequence of the thermodynamic
Maxwell relation: from the identity,
\beq
\frac{\d}{\d B^i} \left(\frac{\d F}{\d \Omega^i}\right) = \frac{\d}{\d \Omega^i} \left(\frac{\d F}{\d B^i}\right)\,,
\eeq
where $F$ is the free energy under rotation and in a magnetic field,
\beq
{\rm d} F = - S {\rm d} T - n_{\rm B} {\rm d} \mu_{\rm B} - {\bm j} \cdot {\rm d} {\bm \Omega} - {\bm m} \cdot {\rm d}{\bm B},
\eeq
we get $\chi_{jB} = \chi_{m \Omega}$ independently of the microscopic details.

In the presence of both ${\bm \Omega}$ and ${\bm B}$, the condition for the emergence of the CSL
solution for finite quark masses is found from the result of ref.~\cite{Brauner:2016pko} as
\beq
\mu_{\rm B} |2\mu_{\rm I}{\bm \Omega} + {\bm B}| \geq 16\pi m_{\pi} f_{\pi}^2\,,
\eeq
which reduces to eq.~(\ref{Omega_CSL}) in the limit of ${\bm B} = {\bm 0}$.

\section{Discussions}
\label{sec:discussion}
Let us discuss the possibility of the realization of the rotation-induced CSL in realistic systems.
To get a rough estimate of $\Omega_{\rm CSL}$ in eq.~(\ref{Omega_CSL}), we consider a rotating nuclear
matter made up from $^{197}_{\ 79}$Au with saturation density $n \approx 0.16 /{\rm fm}^3$, thus
$\mu_{\rm B}^0 \approx 1000$ MeV and $\mu_{\rm I}^0 \approx 10$ MeV.
In this case, $\Omega_{\rm CSL}$ is
\beq
\Omega_{\rm CSL}^0 \approx 6 \times 10^3 \ {\rm MeV},
\eeq
where we take $f_{\pi} \approx 93$ MeV and $m_{\pi} \approx 140$ MeV. Recalling that the recently measured vorticity in heavy-ion collisions averaged over the collision energy from 7.7 GeV to 200 GeV is roughly $10$ MeV~\cite{STAR:2017ckg}, it seems the above $\Omega_{\rm CSL}^0$ is unreachable. However, if the low-energy heavy ion collisions experiments in the future (e.g., the beam energy scan program of the Relativistic Heavy-Ion Collider at BNL, the Nuclotron-based Ion Collider facility at Dubna, and the Facility for Antiproton and Ion Research at Darmstadt) could produce compressed matter with chemical potentials several times larger than $\mu_{\rm B}^0$ and $\mu_{\rm I}^0$, then $\Omega_{\rm CSL}$ would be
\beq
\label{estimate}
\Omega_{\rm CSL}^0 \sim 10^2 \ {\rm MeV},
\eeq
which may possibly be within the experimental reach as the vorticity at lower energy collisions would be higher~\cite{Deng:2016gyh}.
Of course, the medium created by collisions must be cool enough so that the temperature is below the critical temperature of the CSL (which is presumably of the order of QCD scale).

In this paper, we have concentrated on the case at $T=0$ (or small $T$), and investigation of the phase 
diagram at finite $T$ would be desirable in this context. At this moment, we would like to remark that, 
as explicitly shown in ref.~\cite{Brauner:2017mui} for the CSL in a magnetic field, thermal fluctuations 
do not destroy this one-dimensional lattice structure due to the explicit breaking of rotational symmetry 
under global rotation, unlike the situations previously considered, e.g., in refs.~\cite{Hidaka:2015xza, Lee:2015bva}.

One remarkable consequence of the appearance of the chiral soliton lattice is its unusual electromagnetic properties.
As argued in ref.~\cite{Yamamoto:2015maz}, the Wess-Zumino-Witten term $\pi^0 {\bm E} \cdot {\bm B}$
leads to the axion electrodynamics (with $\langle \pi^0 \rangle$ playing the role of the $\theta$ term)
in this QCD matter. As a result, photons become gapped or have gapless nonrelativistic dispersion
depending on their helicity \cite{Yamamoto:2015maz, Ozaki:2016vwu, Qiu:2016hzd, Brauner:2017mui}.
It would be interesting to study possible signatures of such electromagnetic properties in experiments.

Finally, it should be important to theoretically derive the full rotation-induced Wess-Zumino-Witten-type term
without referring to the anomaly matching for the CVE. This is deferred to future work.

\section*{Acknowledgement}
X.~G.~H is supported by the Young 1000 Talents Program of China, NSFC with Grant No.~11535012 and No.~11675041, and
the Scientific Research Foundation of State Education Ministry for Returned Scholars. N.~Y. is supported by JSPS KAKENHI Grant No.~16K17703 and MEXT-Supported Program for the Strategic Research Foundation at Private Universities, ``Topological Science" (Grant No.~S1511006).
N.~Y. thanks the hospitality at Fudan University, where this collaboration was initiated.

\appendix
\section{Two-flavor color superconductivity under rotation}
\label{sec:CSC}
In the main text, we consider QCD with both $\mu_{\rm B}$ and $\mu_{\rm I}$ under rotation.
In this appendix, we consider another example: two-flavor QCD at large $\mu_{\rm B}$ where
two-flavor color superconductor (2SC) is realized. As instanton effects are suppressed at
sufficiently large $\mu_{\rm B}$ \cite{Son:2001jm}, the flavor-singlet $\eta$ meson can also be
regarded as a NG mode (in the chiral limit) associated with the $\U(1)_{\rm A}$ symmetry breaking
by the diquark condensate. In this case, the anomalous term (\ref{L_new}) becomes
\beq
\label{L_eta}
{\cal L}_{\rm anom} =  \frac{1}{f_{\eta}}
\left(\frac{\mu_{\rm B}^2}{\pi^2 N_{\rm c}} + \frac{N_{\rm c} T^2}{3} \right)
{\bm \nabla} \eta  \cdot {\bm \Omega}\,.
\eeq
Note that both of the $\mu$ and $T$ dependent terms above are nonvanishing because
$\Tr[\tau_a] \neq 0$ for a flavor-singlet meson $\eta$ ($a=0$). Note also that, as we argued in
section~\ref{sec:CVE}, the $T$-dependent term can, in principle, get a renormalization and
its coefficient may not be exact unlike the $\mu$-dependent term.

Similarly to the discussion in the main text, this term indicates, e.g., a nonzero angular momentum
density in the presence of $\langle {\bm \nabla} \eta \rangle \neq 0$ as
\beq
{\bm j} = \frac{1}{f_{\eta}}
\left(\frac{\mu_{\rm B}^2}{\pi^2 N_{\rm c}} + \frac{N_{\rm c} T^2}{3} \right)
\langle {\bm \nabla} \eta \rangle \,.
\eeq

\section{Charged pion condensation under rotation}
\label{sec:CPC}
We consider the stability of the CSL state against the charged pion fluctuation. Following ref.~\cite{Brauner:2016pko}, we start with the chiral action
\begin{eqnarray}
\label{chpt}
S=\frac{f_\p^2}{4}\int {\rm d}^4 x\sqrt{-g}\Tr\ls (\nabla_\m\S)^\dag\nabla^\m\S+m_\p^2\lb\S^\dag+\S\rb\rs,
\end{eqnarray}
where $\nabla_\m$ is the covariant derivative, $\nabla_\m\S =\pt_\m\S-i\m_{\rm I}\delta_{0\m}(\t_3\S-\S\t_3)/2$, $g$ is the determinant of $g_{\mu\nu}$ which is the metric tensor of the rotating frame. In matrix form:
\begin{eqnarray}
(g_{\m\n})=
\begin{pmatrix}
1-\O^2 r^2 & y\O & -x\O & 0 \\
y\O & -1 & 0 & 0 \\
-x\O & 0 & -1 & 0 \\
0 & 0 & 0 & -1
\end{pmatrix},
\end{eqnarray}
where $r^2 \equiv x^2 + y^2$.
Parameterize $\S$ as $\S=e^{i\t_3\f} U$ with $U$ a new unitary matrix for which we choose the following form:
\begin{eqnarray}
U=\sqrt{1-\frac{\vec\p^2}{f_\p^2}}+\frac{i\vec\t\cdot\vec\p}{f_\p}.
\end{eqnarray}
Substituting $\S$ and $U$ to eq. (\ref{chpt}) and keeping terms up to second order in $\vec\p$, we obtain
\begin{eqnarray}
\label{chpt3}
S&\approx&\int {\rm d}^4 x\big[ g^{\m\n}\pt_\m \p^+\pt_\n \p^-+ i\pt_z\f\lb \p^+\pt_z \p^--\p^-\pt_z \p^+\rb-ig^{0\m}\m_{\rm I}(\p^+\pt_\m\p^--\p^-\pt_\m\p^+)\non
&&+(\mu_{\rm I}^2-m_\p^2 \cos\f) \p^+\p^-\big]+\int {\rm d}^4 x\ls \frac{1}{2}g^{\m\n}\pt_\m \p_0\pt_\n\p_0+g^{\m\n} f_\p\pt_\m\f\pt_\n\p_0-\frac{1}{2}m_\p^2 \cos\f\p_0\rs,\non
\end{eqnarray}
where $\pi^{\pm}\equiv(\pi_1 \pm i \pi_2)/\sqrt{2}$.
Due to the anomaly matching, an additional action should be added which affects only the $\p_0$ sector:
\begin{eqnarray}
\label{chpt4}
S_{\rm anom}&=&\frac{\m_{\rm B}\m_{\rm I}}{2\p^2 f_\p}\int {\rm d}^4 x \;\O\;
\pt_z\p_0\,.
\end{eqnarray}

We will focus on the charged pions. The dispersion relation of $\p^+$ is given by the following equation,
\begin{eqnarray}
\label{eomcharge1}
\ls(\pt_t-i\O L_z-i\m_{\rm I})^2-\pt_r^2-\frac{1}{r}\pt_r-\frac{\pt_\h^2}{r^2}\rs\p^++ O_z \p^+=0\,,
\end{eqnarray}
where $L_z=-i x\pt_y+iy\pt_x=-i\pt_\h$ and $O_z= -\pt_z^2-2i(\pt_z\f)\pt_z+m_\p^2 e^{-i\f}$. As $L_z$ commutes with other operators, we can write the solution as
\begin{eqnarray}
\p^+=e^{-i\o t}e^{il\h}f(r)g(z),
\end{eqnarray}
with $l$ the orbital angular momentum. Let $\l$ be the eigenvalue of $O_z$. We obtain
\begin{eqnarray}
\ls\pt_r^2+\frac{1}{r}\pt_r-\frac{l^2}{r^2}+(\o+l\O+\m_{\rm I})^2-\l\rs f(r)=0\,.
\end{eqnarray}
The effect of rotation is to introduce a $l$ dependent chemical potential for charged pions. As the system is rotating, it must be finite in the plane transverse to the rotating axis. Assume the system is confined in a cylinder with radius $R$ with the Dirichlet boundary condition,
\begin{eqnarray}
f(R)=0.
\end{eqnarray}
If $(\o+l\O+\m_{\rm I})^2-\l\leq 0$, there is no solution to satisfy the boundary condition. Thus $(\o+l\O+\m_{\rm I})^2-\l>0$. Then the solution is a Bessel function:
\begin{eqnarray}
f(r)=J_l\lb p_\perp r\rb,
\end{eqnarray}
where $p_\perp$ is determined by the boundary condition, $p_\perp=\x_l^n/R$, with $\x_l^n$ the $n$th zero of $J_l(x)$. Thus we have
\begin{eqnarray}
\o=\pm\sqrt{\l+p_\perp^2}-\m_{\rm I}-l\O.
\end{eqnarray}
Let us look at the particle branch corresponding to the plus sign in the above expression. We seek for the condition for pion condensation. If the lowest $\o$ is smaller than zero, the $\p^+$ will condense. This gives (note that $|p_\perp|>|l\O|$ as $\x_l^n>|l|$)
\begin{eqnarray}
\sqrt{\l+(\x_l^1/R)^2}-\m_{\rm I}-l\O<0.
\end{eqnarray}

First, let us consider the chiral limit, $m_\p=0$. In this case, $\f$ is a linear function of $z$ and thus
\begin{eqnarray}
\l=p_z^2+\frac{\O\m_{\rm B}\m_{\rm I}}{\p^2 f_\p^2}p_z=\lb p_z+\frac{\O\m_{\rm B}\m_{\rm I}}{2\p^2 f_\p^2}\rb^2-\frac{(\O\m_{\rm B}\m_{\rm I})^2}{4\p^4 f_\p^4}.
\end{eqnarray}
Thus the condition to have pion condensation is
\begin{eqnarray}
(\x_l^1/R)^2<\frac{(\O\m_{\rm B}\m_{\rm I})^2}{4\p^4 f_\p^4}+(\m_{\rm I}+l\O)^2.
\end{eqnarray}
The above condition shows that if $\m_{\rm I}/(4\p f_\p)$ is sufficiently small ($\m_{\rm I}/(4\p f_\p)$ is required to be small to make the low-energy effective theory applicable), there can not be pion condensation. On the other hand, if $\O=0$, the critical $\m_{\rm I}$ for the pion condensation is given by $\m_{\rm I}>\x_l^1/R$; the rotation thus catalyzes the occurrence of the pion condensation as at finite $\O$ the critical $\m_{\rm I}$ is lowered~\cite{Mameda:2015ria}.

Second, we consider the case away from the chiral limit. We need to solve the eigenvalue problem,
\begin{eqnarray}
\label{eigenoz}
O_z\p^+=\l\p^+,
\end{eqnarray}
with $O_z=-\pt_z^2-2i\pt_z\f\pt_z+m_\p^2 e^{-i\f}$. (What we need is actually the lower bound of $\l$.) This has been done in ref.~\cite{Brauner:2016pko} and one can find that the bottom of $\l$ is given by
\begin{eqnarray}
\min\l=-\frac{m_\p^2}{k^2}\lb2-k^2+\sqrt{1-k^2+k^4}\rb,
\end{eqnarray}
where $k$ is determined through eq. (\ref{condition}).
Therefore, we determine the bottom of $\o$ as
\begin{eqnarray}
\min\o=\min\limits_l\sqrt{\frac{(\x^1_l)^2}{R^2}-\frac{m_\p^2}{k^2}\lb2-k^2+\sqrt{1-k^2+k^4}\rb}-l\O-\m_{\rm I}\,.
\end{eqnarray}
The condition that $\min\o>0$ together with eq. (\ref{Omega_CSL}) determines the parameter region in which the CSL is stable.

\end{document}